\documentclass[conference]{IEEEtran}
\IEEEoverridecommandlockouts
\usepackage{cite}
\usepackage{amsmath,amssymb,amsfonts}
\usepackage{algorithm, algorithmic}
\usepackage{algorithmic}
\usepackage{graphicx}
\usepackage{bm}
\usepackage{textcomp}
\usepackage{xcolor}
\def\BibTeX{{\rm B\kern-.05em{\sc i\kern-.025em b}\kern-.08em
    T\kern-.1667em\lower.7ex\hbox{E}\kern-.125emX}}
\begin{document}

\title{Realization of a Fully Connected Neural Layer Over-the-Air through Multi-hop Amplify-and-Forward Relays\\
\thanks{Identify applicable funding agency here. If none, delete this.}
}

\author{\IEEEauthorblockN{Tolga Girici}
\IEEEauthorblockA{\textit{Dept. of Electrical and Electronics Engineering} \\
\textit{TOBB University of Economics and Technology}\\
Ankara, Türkiye \\
tgirici@etu.edu.tr}
\and
\IEEEauthorblockN{Meng Hua and Deniz Gündüz}
\IEEEauthorblockA{\textit{Dept. of Electrical and Electronics Engineering} \\
\textit{Imperial College London}\\
\textit{United Kingdom} \\
\{m.hua, d.gunduz\}@imperial.ac.uk
}}

\maketitle

\begin{abstract}
We study the problem of implementing a fully-connected layer of a neural network using wireless over-the-air computing. We assume a multi hop system with a multi-antenna transmitter and receiver, along with a number of multi-hop amplify-and-forward relay devices in between. We formulate an optimization problem that optimizes the transmitter precoder, receiver combiner and amplify-and-forward gains, subject to relay device power constraint and transmitter power constraint. We propose an alternating optimization framework that optimizes the imitation accuracy. Simulation study results reveal that multi-hop relaying achieves an almost perfect classification accuracy when used in a neural network. 
\end{abstract}


\section{Introduction}

Modern edge applications (e.g. AR/VR, autonomous systems, massive IoT) need millisecond inference with tight energy budgets and limited backhaul. Digital inference at the cloud creates latency and increases the energy expenditure due to heavy matrix operations. Over-the-air (OTA) computation utilizes the wireless channel itself for the multiply/accumulate operations and avoids quantization, packetization and reconstruction overheads. In the recent literature, OTA computations are  utilized for data fusion, localization, wireless control \cite{csahin2023survey}, federated learning \cite{amiri2020federated},  and neural processing \cite{sanchez2022airnn}. 

Implementing neural layers through OTA computations have been pursued along several paths. The authors in \cite{reus2023airfc} implemented AirFC, an OTA fully-connected neural network layer using a multiantenna transmitter, a single antenna receiver and an OFDM-based transmission. Complementarily, the work in \cite{yang2023over} utilize a MIMO cellular system, wherein each base station (BS) implements one segment of a neural network.  Most recently, the work in \cite{bian2025over} show analog inference over multi-hop MIMO, revealing how cascaded channels can implement deeper computations while exposing new challenges in rank, noise accumulation, and power budgeting across hops.

Convolutional layer of a neural network can also be implemented using OTA computations. For example 1D convolution is implemented in \cite{sanchez2022airnn} while 2D convolution is implemented in \cite{zhang2024radio}. Reconfigurable intelligent surfaces (RISs) are utilized in these implementations. The work in \cite{hua2025implementing} used multiple RISs in order to implement a fully-connected (FC) layer. The effective (cascade) channel is aimed to mimic the weight matrix of a FC layer. Multiple RISs are utilized in order to implement a full-rank effective channel matrix. However, deployment and maintenance of multiple RISs for this purpose requires significant investment.

Amplify-and-forward (AF) relaying has been previously considered in OTA computing in \cite{wan2023energy}, \cite{tang2022node}, \cite{wang2022amplify}. The authors in \cite{wan2023energy} used an AF relay to help transmission from sensors to a fusion center. Sensor node - AF relay association and optimal scheduling was studied in \cite{tang2022node}.  AF relaying can also be used in federated learning \cite{lin2022relay}. Here devices send their local models to the access point (AP), and the models are aggregated OTA. In this scenario,  AF relays can be used to improve the signal quality.  In \cite{wang2022amplify}, the authors envisioned a hierarchical network wherein intermediate AF relays help data fusion with OTA computations. This idea was extended to cognitive radio networks in \cite{yao2024joint}. Finally, authors in \cite{luo2023joint} propose a noise-aided scheme in order to avoid the relayed data being wiretapped by eavesdroppers. Multihop AF relaying was not previously considered in OTA neural network implementation. 

In this work we utilize AF relay devices between a multi-antenna BS and a multi-antenna receiver (Rx) in order to implement/imitate an OTA FC layer. AF relaying has been previously utilized for OTA sensor fusion \cite{wan2023energy}, \cite{tang2022node} but not for implementation of a neural network layer. Firstly, AF relays can be useful if there is no direct BS-Rx link. Secondly, even if there is a direct link, this MIMO channel can be rank deficient (e.g., in the millimeter-wave band). Hence, AF relays can be used to improve the rank of the effective channel, without using multiple RISs. Thirdly, using relays improve the received SNR at the Rx. Multi-hop (rather than two-hop) AF relaying can be useful and necessary when the BS-Rx distance is large. Our analysis clarifies scaling and feasibility with distance/pathloss, and we report extensive simulations showing high imitation accuracy, robustness to noise, and diminishing gains with the number of hops. Proposed multi-hop OTA FC  subsumes two-hop as a special case.

\section{System Model}

\begin{figure}[htbp]
\centerline{\includegraphics[width=\columnwidth]{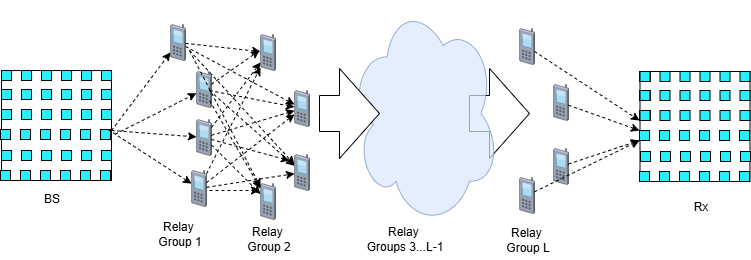}}
\caption{Multi-hop OTA computing system model}
\label{fig:MultiHopAirFC}
\end{figure}

We assume a multi-antenna BS with $N_t$ antennas transmitting to a receiver (Rx) through a number of relay devices, as shown in Fig. \ref{fig:MultiHopAirFC}. There are $K$ single-antenna relay devices randomly distributed over an area,  organized into $L$ relay groups in series; group $l$
has $K_l$ single-antenna AF relays (typically $K_l=K/L$, so $\sum_{l=1}^L K_l=K$). Let $\mathbf{H}_{1} \in \mathbb{C}^{K_1\times N_t}$ denote the complex  baseband channel matrix from the BS to the first relay group, where $[\mathbf{H}_{1}]_{i,j}$ is the channel gain from the $j^{th}$ antenna of the BS to the $i^{th}$ device in group $1$.

Suppose that, upon receiving the signal from the BS, each device  amplifies and forwards this signal to the second relay group. We assume that relay groups access the channel in a TDMA manner. Let $\mathbf{x}\in \mathbb{C}^{N}$ be the transmitted baseband complex signal vector. The symbols are uncorrelated, i.e. $E[\mathbf{x}\mathbf{x}^H]=\mathbf{I}$, and therefore, their transmissions do not interfere with each other.  Let $\mathbf H_{l+1} \in \mathbb C^{K_{l+1}\times K_l}, l=1,\dots,L-1$,  be the channel matrix between relay groups $l$ and $l+1$. Lastly, the $L^{th}$  relay group transmits to a Rx, which has $N$ receive antennas. Let the column vector $\mathbf{g}_k  \in \mathbb{C}^{1\times N}$ be the complex baseband channel from node $k$ to the receiver, which are collected into $\mathbf H_{L+1}=\mathbf{G}=[\mathbf{g}_1,\ldots,\mathbf{g}_K]\in\mathbb{C}^{N_r\times K_L}$. Upon receiving the signal, relay device $k$ in group $l$ amplifies the signal with complex weight $a_k$ and forwards it to the next stage.  Let us define the diagonal forwarding matrix of relay group $l$ as  $\mathbf A_l \triangleq \mathrm{diag}(\mathbf a_l)\in\mathbb C^{K_l\times K_l}$ with
$\mathbf a_l\in\mathbb C^{K_l}$.

We also assume the existence of a direct channel from the BS to the Rx, given by $\mathbf{H}_{0} \in \mathbb{C}^{N\times N_t}$. The effective baseband channel of the system can be defined as follows,

\begin{equation}
  \mathbf H_{\rm eff}
  = \mathbf H_{0} +\mathbf H_{L+1}\mathbf A_L \mathbf H_L \cdots \mathbf A_2 \mathbf H_2 \,\mathbf A_1 \mathbf H_1
  \ \in \ \mathbb C^{N_r\times N_t},
\end{equation}For an input $\mathbf x\in\mathbb C^{N}$, the Rx observation is
\begin{equation}
  \mathbf y \ =\ \mathbf F_2\big( \mathbf H_{\rm eff}\,\mathbf F_1 \mathbf x + \mathbf n_{\rm in}\big),
  \label{eq:inputoutput}
\end{equation}
where $\mathbf n_{\rm in}$ is the cumulative noise at the Rx input, which is derived as follows: Each relay group $l$ introduces  noise $\mathbf n_l\sim\mathcal{CN}(\mathbf 0,\sigma_{u,l}^2\mathbf I_{K_l})$
\emph{before} amplification, while $\mathbf n_c\sim\mathcal{CN}(\mathbf 0,\sigma_c^2\mathbf I_{N_r})$ is added at the Rx. Due to linear operations $\mathbf n_{\rm in}$ is also Gaussian with $\mathcal{CN}(\mathbf 0,\mathbf R_n^{\rm in})$. The aggregate noise covariance at the Rx is, 

\begin{equation}
\label{eq:Rn_in}
  \mathbf R_n^{\rm in}
  \ =\ \sigma_c^2\,\mathbf I_N \ + \ \sum_{j=1}^{L}\ \mathbf T_j\,\sigma_{u,l}^2\,\mathbf T_j^{H}.
\end{equation} Here $\mathbf T_j$ define the transfer matrix from group $j$ to the Rx input:
\begin{multline}
  \mathbf T_j \ \triangleq\ \mathbf H_{L+1}\mathbf A_L \mathbf H_L \cdots \mathbf H_{j+1}\mathbf A_j
  \ \in\ \mathbb C^{N_r\times K_j},
  \qquad \\j=1,\dots,L.
\end{multline}

\subsection{Multi-Hop AirFC with AF}
Let $\mathbf W\in\mathbb C^{N\times N}$ be the target weight matrix of the FC layer. We want to emulate this digital FC layer. That is, we want the input-output relation in  (\ref{eq:inputoutput}) to resemble the behavior of the FC layer:  

\begin{equation}
  \label{eq:fc}
  \mathbf{y} = \mathbf{W}\mathbf{x} + \mathbf{b},\qquad
  \mathbf{x},\mathbf{y}\in\mathbb{C}^{N\times 1},\; \mathbf{W}\in\mathbb{C}^{N\times N}.
\end{equation}where the symbols $\mathbf{x}$ are assumed to be uncorrelated and normalized, i.e. $\mathbb{E}[\mathbf{x}\mathbf{x}^H]=\mathbf{I}_N$. A BS with $N_t$ antennas precodes $\mathbf{x}$ using precoding matrix $\mathbf{F}_1\in\mathbb{C}^{N_t\times N}$  and the relay devices in group $l$ amplify-and-forward their received signals using gain matrix $\mathbf{A_l}$. Then, the received signal from the $L^{th}$ relay group is multiplied by a complex combining matrix  $\mathbf{F}_2 \in \mathbb{C}^{N_r\times N }$ at the receiver. The output signal becomes,

\begin{equation}
  \label{eq:r}
  \mathbf{y} = \mathbf{F}_2\mathbf{H}_{\rm eff}\mathbf{F}_1\mathbf{x} + \mathbf{F}_2\mathbf{n}_{in}.
\end{equation}


\subsection{Imitation Objective and Constraints}

We seek OTA parameters $(\mathbf F_1,\mathbf F_2,\{\mathbf A_l\})$ that minimize
an imitation error plus a noise penalty:
\begin{subequations}
\label{eq:airfc_problem}
\begin{align}
  \min_{\mathbf F_1,\mathbf F_2,\{\mathbf A_l\}}\
  & \underbrace{\big\|\,\mathbf F_2 \mathbf H_{\rm eff} \mathbf F_1 - \mathbf W\,\big\|_F^2}_{\text{FC imitation error}}
    \ +\ \underbrace{\mathrm{tr}\!\big(\mathbf F_2 \mathbf R_n^{\rm in} \mathbf F_2^H\big)}_{\text{noise propagation}}
  \label{eq:main_problem}\\
  \text{s.t.}\quad
  & \|\mathbf F_1\|_F^2 \ \le\ P_{\max}, \label{eq:tx_power}
  \\
  & \mathbb E\!\left[\,|a_{l,k} u_{l,k}|^2\,\right]\ \le\ P_{l,k}, \nonumber\\
  & \quad  \quad \quad \forall\,l=1,\dots,L,\ \ k=1,\dots,K_l. \label{eq:relay_power}
\end{align}
\end{subequations}
In \eqref{eq:relay_power}, $u_{l,k}$ denotes the (complex) signal incident on relay $(l,k)$ before
amplification. A conservative instantaneous surrogate for its variance is
\begin{equation}
  \label{eq:relay_input_power}
  p^{\rm in}_{l,k}\ \approx\
  \left\|\big(\mathbf H_l \mathbf A_{l-1}\mathbf H_{l-1}\cdots \mathbf A_1 \mathbf H_1 \mathbf F_1\big)_{k,:}\right\|_2^2
  \ +\ \sigma_{u,l}^2,
\end{equation}
so that $|a_{l,k}|^2\,p^{\rm in}_{l,k}\le P_{l,k}$.

We add a noise penalty to the objective because even if the effective channel $\mathbf F_2 \mathbf H_{\rm eff} \mathbf F_1$ resembles $\mathbf W$ noise can be so much amplified the the Rx that the useful signal is buried in noise. 

\paragraph{Rank Considerations}: The realizable map $\mathbf H_{\rm chain}\triangleq\mathbf{F}_2\mathbf{H}_{\rm eff}\mathbf{F}_1$ is ultimately limited by per-hop bottlenecks, such as the matrix rank: 
\begin{equation}
  \mathrm{rank}(\mathbf H_{\rm chain}) \ \le\ 
  \min\!\big\{\, N_t,\ K_1,\dots,K_L,\ \mathrm{rank}(\mathbf H_\ell)\ \forall \ell \,\big\}.
\end{equation} Assuming $N_t=N_r=N$, $\mathbf H_{\rm chain}$ and weight matrix $\mathbf{W}$ are square. Please note that achieving an $N\times N$ FC imitation with negligible error does not require $\mathbf H_{\rm chain}$ to be full rank. After all $\mathbf{W}$ itself can be rank-deficient. Nevertheless we can say that the condition $\mathrm{rank}(\mathbf H_{\rm chain})\geq \mathrm{rank}(\mathbf W)$ is required for a good imitation accuracy.

\subsection{CSI Acquisition and Synchronization}
The proposed design assumes perfect CSI of all inter-hop channels \(\{\mathbf H_\ell\}_{\ell=0}^{L+1}\), which can be obtained via pilot-based training. Under the TDMA operation assumed across relay groups, channel estimation can be performed sequentially per hop using orthogonal pilots. The obtained channel information is collected at the BS, which centrally performs the optimization.

OTA neural network implementation relies also on coherent signal superposition at each node, which requires tight time, frequency, and phase synchronization across transmitting nodes. Practical implementations require pilot-aided synchronization or relay-level re-synchronization mechanisms to mitigate the synchronization errors that may accumulate across hops.

\section{Alternating Optimization (AO) Based Solution}
We will solve the optimization problem in (\ref{eq:airfc_problem}), (\ref{eq:tx_power}), (\ref{eq:relay_power}) using AO, where we iteratively optimize with respect to BS precoder $\mathbf{F}_1$, Rx combiner $\mathbf{F}_2$ and relay AF gains $\mathbf{A}_l, l=1,\ldots,L$. 
Recall the hop channels
\(
\mathbf H_1\in\mathbb C^{K_1\times N},\,
\mathbf H_{l+1}\in\mathbb C^{K_{l+1}\times K_l}\ (l=1,\dots,L-1),\,
\mathbf H_{L+1}\in\mathbb C^{N\times K_L}
\),
optionally a direct link \(\mathbf H_0\in\mathbb C^{N\times N}\),

\subsection{Block 1: BS Precoder \(\mathbf F_1\)} We first
fix \((\mathbf F_2,\{\mathbf A_l\})\) and denote
\(\mathbf \Xi \triangleq \mathbf F_2 \mathbf H_{\rm eff}\in\mathbb C^{N\times N}\).
Ignoring constants, the subproblem is
\begin{equation}
\min_{\mathbf F_1}\ \big\|\mathbf \Xi \mathbf F_1 - \mathbf W\big\|_F^2
\quad\text{s.t.}\ \|\mathbf F_1\|_F^2 \le P_{\max}.
\end{equation} We set the gradient \(\partial/\partial \mathbf F_1^*\) of
\(\|\mathbf \Xi \mathbf F_1 - \mathbf W\|_F^2 + \lambda \|\mathbf F_1\|_F^2\) to zero. Here $\lambda $ is the Lagrange multiplier. From the equality 
\(
\mathbf \Xi^H(\mathbf \Xi \mathbf F_1 - \mathbf W) + \lambda \mathbf F_1 = \mathbf 0
\), optimal $\mathbf F_1(\lambda)$ is found as,
\begin{equation}
\label{eq:F1_opt}
\mathbf F_1(\lambda) = \big(\mathbf \Xi^H \mathbf \Xi + \lambda \mathbf I\big)^{-1}\mathbf \Xi^H \mathbf W,\quad \lambda\ge 0.
\end{equation}
If the unconstrained solution \(\lambda=0\) violates \eqref{eq:tx_power}, choose \(\lambda\) (e.g., by bisection) so that \(\|\mathbf F_1(\lambda)\|_F^2=P_{\max}\) is satisfied.
Since \(\|\mathbf F_1(\lambda)\|_F^2\) is strictly decreasing in \(\lambda\), bisection converges.

\subsection{Block 2: Rx Combiner \(\mathbf F_2\)}
This time , we fix \((\mathbf F_1,\{\mathbf A_l\})\) and define
\(\mathbf U \triangleq \mathbf H_{\rm eff}\mathbf F_1\) and \(\mathbf R_n\triangleq \mathbf R_n^{\rm in}\) from \eqref{eq:Rn_in}.
The subproblem becomes,
\begin{equation}
\min_{\mathbf F_2}\ \big\|\mathbf F_2 \mathbf U - \mathbf W\big\|_F^2 + \mathrm{tr}\big(\mathbf F_2 \mathbf R_n \mathbf F_2^H\big).
\end{equation} We set \(\partial/\partial \mathbf F_2^*\) of the objective to zero and obtain the equality
\(
(\mathbf F_2 \mathbf U - \mathbf W)\mathbf U^H + \mathbf F_2 \mathbf R_n = \mathbf 0
\),
and right-multiply by \((\mathbf U \mathbf U^H + \mathbf R_n)^{-1}\).

The unique minimizer is
\begin{equation}
\label{eq:F2_opt}
\mathbf F_2^\star = \mathbf W\,\mathbf U^H\,\big(\mathbf U \mathbf U^H + \mathbf R_n\big)^{-1}.
\end{equation}

\subsection{Block 3: Relay Gains \(\mathbf A_l=\mathrm{diag}(\mathbf a_l)\)}
Fix \((\mathbf F_1,\mathbf F_2,\{\mathbf A_i\}_{i\neq l})\).
We fold the chain to isolate \(\mathbf A_l\) by defining the following pre- and post-product matrices,
\begin{align}
\mathbf U_l &\triangleq \mathbf F_2 \,\mathbf H_{L+1}\mathbf A_L \mathbf H_L \cdots \mathbf A_{l+1}\mathbf H_{l+1}
\ \in \ \mathbb C^{N\times K_l},\\
\mathbf V_l &\triangleq \mathbf H_{l}\,\mathbf A_{l-1}\mathbf H_{l-1}\cdots \mathbf A_{1}\mathbf H_{1}\,\mathbf F_1
\ \in\ \mathbb C^{K_l\times N}.
\end{align}
The current effective map at the detector is \(\mathbf M = \mathbf F_2 \mathbf H_{\rm eff} \mathbf F_1=\sum_{j=1}^{L-1}\mathbf U_j \mathrm{diag}(\mathbf a_j)\mathbf V_j+\mathbf{F}_2\mathbf{H}_0\mathbf{F}_1\).
We subtract the present layer-\(l\) contribution to form the  ''residual target" (i.e. the component that does not depend on $\mathbf a_l$ )
\begin{equation}
\label{eq:El}
\mathbf E_l \triangleq \mathbf W - \big(\mathbf M - \mathbf U_l \mathrm{diag}(\mathbf a_l)\mathbf V_l\big).
\end{equation}
We solve the following least squares optimization problem with regularization:
\begin{equation}
\label{eq:al_subprob}
\min_{\mathbf a_l}\ \big\|\mathbf U_l \mathrm{diag}(\mathbf a_l)\mathbf V_l - \mathbf E_l\big\|_F^2
\;+\; \mathbf a_l^H \mathbf D_l \mathbf a_l,
\end{equation}
with \(\mathbf D_l\succeq \mathbf 0\) an optional (noise-aware) diagonal regularizer.

\paragraph*{Vectorization and Khatri--Rao}
Using \(\mathrm{vec}(\mathbf U\,\mathrm{diag}(\mathbf a)\,\mathbf V) = (\mathbf V^T \odot \mathbf U)\,\mathbf a\),
\eqref{eq:al_subprob} becomes,
\begin{equation}
\min_{\mathbf a_l}\ \big\|\underbrace{(\mathbf V_l^T \odot \mathbf U_l)}_{\mathbf B_l}\mathbf a_l - \underbrace{\mathrm{vec}(\mathbf E_l)}_{\mathbf e_l}\big\|_2^2
\;+\; \mathbf a_l^H \mathbf D_l \mathbf a_l.
\end{equation}where \(\odot\) denotes the Khatri--Rao product. The normal equations are

\begin{equation}
\label{eq:al_normal}
\big(\mathbf B_l^H \mathbf B_l + \mathbf D_l\big)\mathbf a_l = \mathbf B_l^H \mathbf e_l.
\end{equation}
Using the Gram identity \((\mathbf V^T \odot \mathbf U)^H(\mathbf V^T \odot \mathbf U)=(\mathbf V\mathbf V^H)\odot(\mathbf U^H\mathbf U)\),
we obtain

\begin{align}
\bm\Gamma_l &\triangleq (\mathbf V_l \mathbf V_l^H)\odot(\mathbf U_l^H \mathbf U_l),\\
\bm\eta_l   &\triangleq \big(\mathbf V_l^T \odot \mathbf U_l\big)^H \mathrm{vec}(\mathbf E_l),
\end{align}
and the closed-form

\begin{equation}
\label{eq:al_solution}
\mathbf a_l^\star = \big(\bm\Gamma_l + \mathbf D_l\big)^{-1}\bm\eta_l.
\end{equation}

\paragraph*{Noise-aware regularizer.} The matrix $\mathbf B_l^H \mathbf B_l$ is frequently ill-conditioned (e.g., line-of-sight (LoS)/low-rank hops, correlated beams), which makes the unconstrained solution noise-sensitive and yields exploding relay gains (even if we project onto per-relay power constraints afterwards). A practical diagonal choice is

\begin{equation}
\label{eq:Dl}
\mathbf D_l = \sigma_{u,l}^2\,\mathrm{diag}\!\big(\mathrm{diag}(\mathbf U_l^H \mathbf U_l)\big),
\end{equation}which reflects the relay-injected noise and the downstream gain. It treats each relay with a different “effective” regularization, proportional to how much its noise will be seen at the receiver.

\paragraph*{Per-relay power projection.}
After \eqref{eq:al_solution}, we enforce \eqref{eq:relay_power} via magnitude projection using
the surrogate \eqref{eq:relay_input_power}:
\begin{equation}
\label{eq:proj_power}
a_{l,k} \ \leftarrow\ 
\min\!\left\{|a_{l,k}|,\ \frac{\sqrt{P_{l,k}}}{\sqrt{p^{\rm in}_{l,k}}}\right\}
\,\frac{a_{l,k}}{|a_{l,k}|},\quad k=1,\dots,K_l.
\end{equation}

\subsection{Algorithm and Convergence}

Algorithm \ref{alg:algorithm} shows the pseudocode of the proposed AO algorithm. Line 1 shows the inputs, which are the digital baseline weight matrix of the FC layer, multihop channel matrices, noise powers and power budgets. Line 2 initializes the algorithm. Lines 3-13 constitute the main loop, which iterates until convergence. Line 4 builds the effective cascade channel matrix $\mathbf H_{\rm eff}$. Lines 5 and 6 update $\mathbf F_1$ and $\mathbf F_2$, respectively. Lines 7-11 perform relay power optimization and projection for each relay group.  Implementing this algorithm requires channel state information (CSI) of all the links in the system. Effects of imperfect CSI and other physical layer imperfections are beyond the scope of this paper. 
\begin{algorithm}[t]
\caption{AO for Multihop AF AirFC}
\begin{algorithmic}[1]
\label{alg:algorithm}
\STATE \textbf{Input:} \(\mathbf W\), channels \(\{\mathbf H_\ell\}_{\ell=0}^{L+1}\), noise powers, power budgets.
\STATE \textbf{Init:} \(\mathbf F_1=\mathbf I\), \(\mathbf F_2=\mathbf I\);
\(\mathbf a_l=\rho\,\mathbf 1\) (small \(\rho\)) projected by \eqref{eq:proj_power}.
\REPEAT
\STATE Build \(\mathbf H_{\rm eff}=\mathbf H_0+\mathbf H_{L+1}\mathbf A_L\mathbf H_L\cdots \mathbf A_1\mathbf H_1\).
\STATE \textbf{Update \(\mathbf F_1\):} Form \(\mathbf \Xi=\mathbf F_2\mathbf H_{\rm eff}\). Compute \(\mathbf F_1(\lambda)\) by \eqref{eq:F1_opt}, choose \(\lambda\) by bisection to satisfy \(\|\mathbf F_1\|_F^2\le P_{\max}\).
\STATE \textbf{Update \(\mathbf F_2\):} Form \(\mathbf U=\mathbf H_{\rm eff}\mathbf F_1\), build \(\mathbf R_n^{\rm in}\) via \eqref{eq:Rn_in}, then set \(\mathbf F_2\leftarrow \mathbf W \mathbf U^H(\mathbf U\mathbf U^H+\mathbf R_n^{\rm in})^{-1}\).
\FOR{\(l=1\) to \(L\)}
\STATE Build \(\mathbf U_l,\mathbf V_l\) (cached prefixes/suffixes), compute \(\mathbf E_l\) by \eqref{eq:El}.
\STATE Solve \(\mathbf a_l\leftarrow (\bm\Gamma_l+\mathbf D_l)^{-1}\bm\eta_l\) with \(\bm\Gamma_l,\bm\eta_l\) as above.
\STATE Project \(\mathbf a_l\) by \eqref{eq:proj_power}.
\ENDFOR
\STATE Evaluate objective in \eqref{eq:main_problem}.
\UNTIL{relative decrease \(<\) tolerance or max iterations.}
\end{algorithmic}
\end{algorithm}

Each block subproblem is convex with a unique global minimizer given the other blocks fixed, hence the AO produces a non-increasing objective sequence and converges to a stationary point of \eqref{eq:main_problem}.

The proposed AO algorithm is executed once before an inference session under a block fading assumption. We assume that channel is coherent during the session. Before the session, pilot transmission is first performed to estimate all hop channels. CSI obtained by all nodes are shared with the BS, and the BS performs the AO-based parameter optimization in a centralized manner. The resulting parameters $\mathbf F_1, \mathbf F_2, \mathbf A_l, l=0,...,L$  are then shared with the relay devices and used during the signal transmission phase.

\section{Numerical Results}
In this section, we present numerical results to evaluate the performance of the proposed multi-hop AirFC framework. We consider the simulation parameters in Table \ref{tab:simparam}. Consider a rectangular service area of size $D_{\max}\times D_{\max}$ meters, partitioned into $L$ rectangular regions of size $D_{\max}\times D_{\max}/L$ arranged serially between a multi-antenna BS and a multi-antenna Rx. Each small cell hosts a \emph{group} of $\tfrac{K}{L}$ single-antenna AF relays. We assume the 3GPP UMi Street Canyon pathloss model in the BS-RX, BS-Relay and Relay-Rx links. Pathloss between the relay devices are modeled according to the TR 38.901 sidelink channel model. LoS probability of any link depends on the link distance according to the model. 

The complete neural network architecture comprises the following layers: one input layer, one convolutional (Conv) layer ($2$ output channels with a kernel size of $3$, stride of $4$, and padding of $1$), real-to-complex (R2C) layer (converts the real-valued input into a complex-valued output with half the original dimensionality), one complex FC layer with complex ReLU activation,  complex batch normalization layer, power normalization layer, precoder, multihop AF relaying, combiner, complex ReLU activation, complex-to-real (C2R) layer (that converts a complex-valued input into a real-valued output with twice the dimensionality.), a real FC layer and an output. 

In the plots given below, each point on the plot is an average of $20$ channel realizations. Plots also involve error bars that show deviations around the mean accuracy. In each plot the dashed straight black line denotes the baseline digital accuracy. Proposed multihop OTA FC layer (denoted by \emph{AirFC} in the plots) aims to achieve this baseline. Digital baseline is created by training on the Fashion-MNIST dataset \cite{hua2025implementing}. Accuracy of the digital baseline is $84.5\%$.

\begin{table}[htb]
    \centering
    \begin{tabular}{|c|c|}
    \hline
        Parameter & Value \\\hline
        Relay devices/group & $K=6,12,..., 54, 60$  \\\hline
        Number relay clusters & $L=5$  \\\hline        
        Number of antennas  &  $N=N_t=N_r=49$ \\\hline 
        BS Power  & $P_{max}=N$ W\\\hline
        Relay Power & $P_{k}=0.1, 1$ W \\\hline
        Carrier frequency & $f_c=28$ GHz \\\hline
        Noise p.s.d. & $N_o=-174$dBm \\\hline
        Bandwidth & $B=300$MHz  \\\hline
        BS,Rx height & $5$ meters\\\hline
        AF relay height & $1.5$ meters\\\hline
        Network diameter & $D_{max}=100, 200$m \\\hline
        Pathloss & 3GPP UMi Street Canyon (NLoS)\\\hline
        BS-Rx MIMO Channel & Ricean ($\kappa=0,$ dB) \\\hline
        BS-device and device-Rx channel & Rich Scattering \\\hline
        Noise power & $\sigma_{u,l}^2=\sigma_c^2=N_oB$\\\hline
    \end{tabular}
    \caption{Simulation Parameters}
    \label{tab:simparam}
\end{table}

\begin{figure}[htb]
\centerline{\includegraphics[width=\columnwidth]{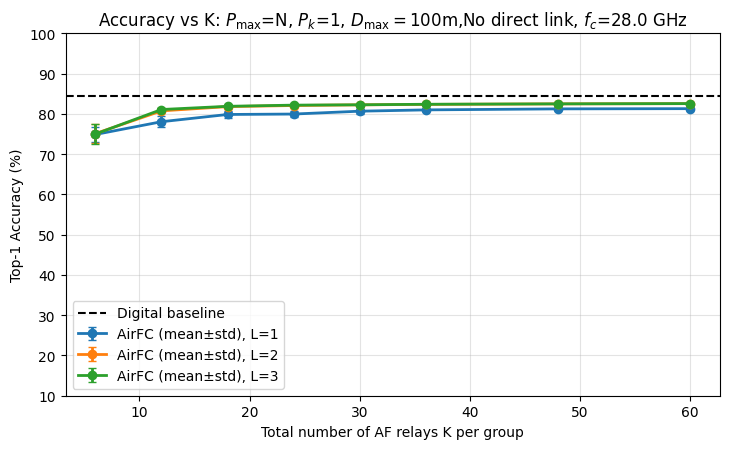}}
\caption{Accuracy vs. Number of Relay Devices per group ($K$). Bs-Rx link blocked, $D_{max}=100$m, AF relay power $P_k=1$ W}
\label{fig:AccvsKPk1Dmax100NDL}
\end{figure}

Fig. \ref{fig:AccvsKPk1Dmax100NDL} shows the classification accuracy vs number of relay devices per group ($K$) for relay power $P_k=1$ W, network size $D_{max}=100$m, and number of groups $L=1,2,3$. Direct BS-Rx link is assumed to be blocked.  For this small-sized area, distances are short and the results reveal that accuracy is very close to the digital baseline for all $L=1,2,3$ groups and for AF relays per group ($K$) greater than $12$. Error bars in this plot are very small, which shows that our multihop imitation framework is very stable. In this case a larger $L$ does not provide a significant improvement in accuracy, because the accuracy is already close to the digital baseline.

\begin{figure}[htb]
\centerline{\includegraphics[width=\columnwidth]{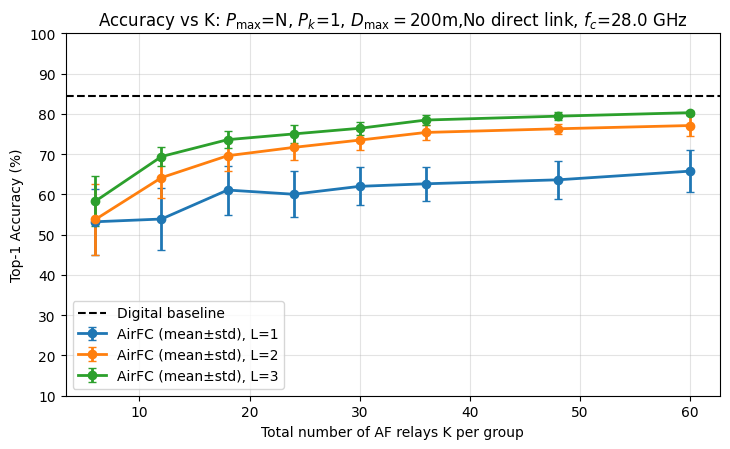}}
\caption{Accuracy vs. Number of Relay Devices per group ($K$). Bs-Rx link blocked, $D_{max}=200$m, AF relay power $P_k=1$ W}
\label{fig:AccvsKPk1Dmax200NDL}
\end{figure}

Figure \ref{fig:AccvsKPk1Dmax200NDL} shows the classification accuracy vs number of relay devices \emph{per group} for $P_k=1$ W and a larger area of $D_{max}=200$m. Again, direct BS-Rx link is assumed to be blocked.   In this case we can see the divergence among different number of groups $L=1,2$ and $3$. Using a single relay group ($L=1$) provides a poor accuracy. $L=2$ significantly improves the imitation accuracy. A larger $L$ achieves a higher accuracy, because the transmission link distances become shorter and noise buildup at the Rx decreases.  Further increase of $L$ brings a diminishing return. There are two reasons for this. First of all the $L=3$ performance is already very close to the digital baseline. Secondly, multihop AF relaying builds up noise at each new hop, which leaves less room for the amplifier gain $a_k$. Variance of the performance for $L=1$ is high since the AF relays are distributed over a large area. $L=2$ and $3$ provide more stable performances.

\begin{figure}[htbp]
\centerline{\includegraphics[width=\columnwidth]{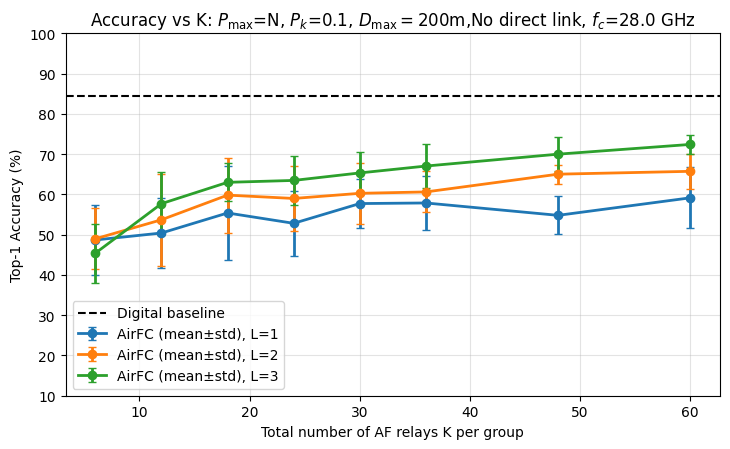}}
\caption{Smaller relay power: Accuracy vs. Number of Relay Devices per group ($K$). Bs-Rx link blocked, $D_{max}=200$m, AF relay power $P_k=0.1$ W}
\label{fig:AccvsKPk01Dmax200NDL}
\end{figure}

Fig. \ref{fig:AccvsKPk01Dmax200NDL} shows the classification accuracy for a smaller AF relay power, $P_k=0.1$ W. Results reveal that imitation accuracy improves with increasing the number of groups $L$ from $1$ to $3$. As expected the accuracy performance is worse than that of $P_k=1$W. Moreover, the variance is higher.

\begin{figure}[htbp]
\centerline{\includegraphics[width=\columnwidth]{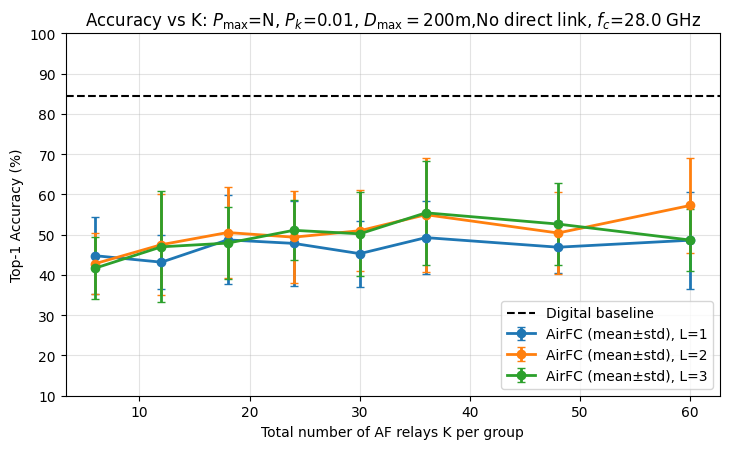}}
\caption{Even smaller relay power: Accuracy vs. Number of Relay Devices per group ($K$). Bs-Rx link blocked, $D_{max}=200$m, AF relay power $P_k=0.01$ W}
\label{fig:AccvsKPk001Dmax200NDL}
\end{figure}

In Fig. \ref{fig:AccvsKPk001Dmax200NDL} we show the classification accuracy for an even smaller AF relay power budget of $0.01$ W. In this case, the performance is poor and very volatile for all $L$ and $K$ values. Results reveal that for a good imitation accuracy sufficient relay power is needed. 

The results in Fig. \ref{fig:AccvsKPk01Dmax200NDL} and Fig. \ref{fig:AccvsKPk001Dmax200NDL} highlight the impact of relay power on imitation accuracy. While increasing relay power improves accuracy, it also increases total energy consumption across the network. Multi-hop relaying introduces a trade-off: shorter transmission distances reduce path loss, but relay amplification leads to higher total energy usage. Future work should consider energy-efficient deployments that strike a trade-off between accuracy and energy-efficiency.

\begin{figure}[htbp]
\centerline{\includegraphics[width=\columnwidth]{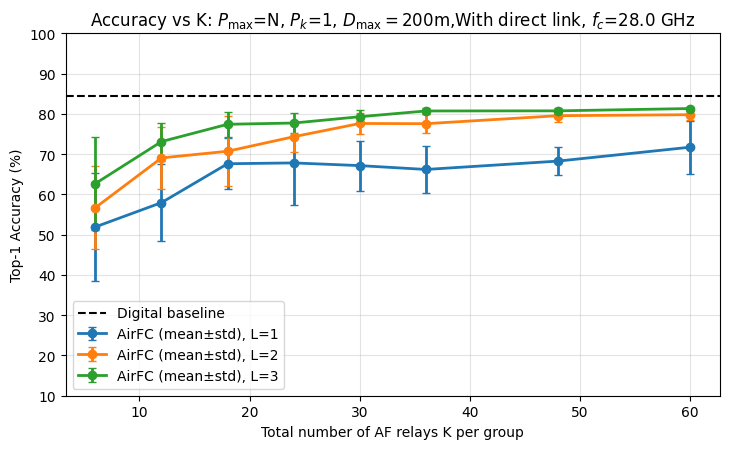}}
\caption{Direct link enabled: Accuracy vs. Number of Relay Devices per group ($K$). Bs-Rx link $\kappa=0$ dB, $D_{max}=200$m, AF relay power $P_k=1$ W}
\label{fig:AccvsKPk1Dmax200WDL}
\end{figure}

Lastly, in Fig. \ref{fig:AccvsKPk1Dmax200WDL}, we present the imitation accuracy in the presence of a direct BS-Rx link for $D_{max}=200$m, and AF relay power of $P_k=1$W. When compared with Fig. \ref{fig:AccvsKPk1Dmax200NDL} direct link slightly improves the performance. However, it can be said that a great part of the imitation accuracy is due to the multi-hop AF relay system. The direct link provides only an extra $1\%$ improvement. Results also reveal that even $K=20$ relays per group is sufficient for a decent performance. 

{\subsection{Signaling and Computational Overhead} The proposed AO algorithm requires iterative updates involving matrix inversions and per-hop computations. The computational complexity scales with the number of antennas and relay nodes, while the signaling overhead depends on the availability and exchange of CSI across the network. However, these operations can be executed at the BS or edge infrastructure, without creating additional burden on end devices}

\section{Conclusions}

In this work we studied the problem of OTA implementation of a FC neural network layer using MIMO transmitter, receiver and a number of AF relays in between. The relays are organized in a series of relay groups in order to achieve multi-hop transmission and overcome wireless pathloss. We proposed an AO method to optimize the MIMO precoder, combiner and AF-relay gains.  Case-study results reveal that close-to-optimal classification accuracy can be obtained using two or three relay groups, even when the direct link is blocked. 

This work relies on some idealistic assumptions, such as perfect CSI and synchronization. Future work may involve investigating the effects of synchronization and channel estimation errors among relays and methods to mitigate them. Moreover, multihop relaying introduces delay and energy expenditure. It is a future work to study the energy efficiency of the relays, selection of an optimal set of relays when the number of relays is large and strike a good balance between energy expenditure and accuracy. Delay is also an important issue. Delay increases due to multihop transmissions and training overhead. However,  improved link quality may also reduce retransmissions and improve accuracy. This trade-off is important for latency-sensitive edge inference applications. Besides these, energy harvesting relay devices and transmission of information concurrently with OTA computing are promising directions for future research.


\bibliographystyle{IEEEtran}
\bibliography{OTAC}

\end{document}